\documentclass[12pt]{iopart}
\usepackage{iopams}
\usepackage{graphicx}
\usepackage{cite}
\newlength{\minitwocolumn}
\newlength{\figsmall}
\newlength{\figwidth}
\newlength{\figlarge}
\begin{document}

\title[Magnetic properties of a Kondo insulator
with RKKY interaction
]
{Magnetic properties of a Kondo insulator
with RKKY interaction: \\
Extended dynamical mean field study}

\author{Takuma Ohashi\footnote[3]
{ohashi@tp.ap.eng.osaka-u.ac.jp}, 
Sei-ichiro Suga, 
and Norio Kawakami 
}

\address{Department of Applied Physics, Osaka University, 
Suita, Osaka 565-0871, Japan}

\begin{abstract}
We study the Kondo lattice model 
with the Heisenberg-type RKKY-exchange coupling among localized
 $f$-spins in the presence of a magnetic field. 
By means of an extended dynamical mean field theory 
combined with the non-crossing approximation, 
we investigate the one-particle spectral function and the
dynamical spin correlation function in the Kondo insulating phase.
It is shown that the magnetic field and 
the RKKY exchange interaction both cause the instability to
the antiferromagnetic order with enhanced transverse
 spin fluctuations, which give rise to
 a strong renormalization of quasi-particles
 as the system approaches the quantum critical 
point. This leads to a tendency to retain the Kondo insulating gap
up to rather large fields.
\end{abstract}




\section{Introduction}

Electron correlations caused by the interactions between 
conduction electrons and localized magnetic moments 
have provided a variety of important 
subjects in condensed matter physics. 
In dilute magnetic alloys, a local moment 
is screened by conduction electrons via the Kondo effect 
below its characteristic temperature, and the system 
is described as a local Fermi-liquid \cite{hewson}.
In concentrated systems, such as heavy-fermion systems,
 the indirect exchange among the localized moments 
mediated by the RKKY interaction becomes 
important \cite{varma,doniach}.
The interplay between the Kondo effect and the RKKY interaction 
gives rise to competing phases in heavy fermion systems. 

Recent experimental observations of quantum critical behavior
 in heavy fermion materials 
have attracted renewed interest in this field.
In ${\rm CeCu_{6-x}Au_x}$ \cite{lohneysen} and 
${\rm YbRh_2Si_2}$ \cite{trovarelli}, the system 
is driven to the antiferromagnetic phase
by alloying, applying pressure, etc. 
In the vicinity of the quantum critical point, both Kondo and 
RKKY interactions become essential and 
physical properties observed in this region differ 
from those of ordinary Fermi-liquids \cite{stewart}.
Experimental observations of such 
non-Fermi-liquid behavior \cite{schroder} 
stimulated detailed theoretical study beyond 
the standard framework \cite{hertz,millis,moriya,kambe} 
of the quantum critical point \cite{coleman}.

The interplay between Kondo and RKKY interactions 
have been investigated using various methods such as 
slave-particle mean field theory \cite{riseborough,saito},
large-$N$ expansion \cite{newns}, 
quantum Monte Carlo (QMC) simulations \cite{capponi}, 
dynamical mean field theory 
(DMFT) \cite{dmft,jarrell,rozenberg,imai_pam}, etc. 
Recently an extension of the dynamical mean field theory 
has been developed \cite{sachdev,kajueter,smith,chitra},
which allows us to incorporate spatially extended spin 
correlations.
Si {\it et al.} have addressed the problem of non-Fermi-liquid 
behavior near the quantum critical point 
by the extended-DMFT (EDMFT) analysis of 
the Kondo lattice model (KLM) with 
the Ising-type RKKY exchange among 
localized moments \cite{si_qcp,zhu}. 
These investigations focused on magnetic properties around the
quantum critical point in heavy-fermion systems. 
Also, P. Sun and G. Kotliar have addressed the same issues by exploiting 
the periodic Anderson model (PAM) \cite{sun_qcp,sun_pam}. 

Magnetic fields should play a particularly important role if
 the system is in the proximity of magnetic instability, because
the applied fields may possibly trigger a magnetic phase transition. 
Recent studies on the half-filled KLM in two-dimensions 
based on mean field theory and QMC simulations have indeed found that
the magnetic field induces a second-order phase transition 
from the paramagnetic to the transverse 
antiferromagnetic phase \cite{beach,milat}. 
Also, in our previous paper we have investigated the field-induced phase 
transition in the three-dimensional case \cite{our_paper}. 
We have studied the half-filled PAM by means of DMFT combined with QMC.
However, DMFT cannot incorporate inter-site fluctuations due to 
the RKKY-interaction. It is 
desirable to investigate the effects of the RKKY interaction
on electronic properties by means of EDMFT. 

In this paper, we study the KLM  
with the Heisenberg-type RKKY-exchange among localized $f$-spins.
By means of EDMFT combined with the non-crossing approximation 
(NCA) \cite{keiter,bickers,schork,haule}, 
the one-particle spectral function and the dynamical spin susceptibility
are calculated. 
We also investigate the field-induced antiferromagnetic
instability and discuss the remarkable behavior in 
spectral properties in the vicinity of 
the critical point. 

The paper is organized as follows. In the next section we 
briefly summarize our model and method. In section. 3, we present our 
numerical results, and discuss the effects of the RKKY interaction 
in a magnetic field. A brief summary is given in section 4.

\section{Model and Method}

\subsection{Kondo lattice model}

We study the following Kondo lattice model including
the Heisenberg-type RKKY interaction, 
\begin{eqnarray}
H  = 
  &-&t \sum_{\left < i,j \right >,\sigma} 
     c_{i\sigma}^\dag c_{j\sigma}
   - \mu \sum_i c_{i\sigma}^\dag c_{i\sigma}
   + J_K \sum_{i} \mathbf{S}^c_i \cdot \mathbf{S}^f_i
  \nonumber \\
  &+&\frac{J}{2} \sum_{\left < i,j \right >}
     \mathbf{S}^f_i \cdot \mathbf{S}^f_j
   - g \mu_B B \sum _i \left [ S^c_{iz} + S^f_{iz} \right ] ,
\end{eqnarray}
where $c_{i\sigma}$ ($c^\dag_{i\sigma}$) annihilates (creates) 
a conduction electron on the $i$th site with spin $\sigma$, 
$\mathbf{S}_{ci}$ ($\mathbf{S}_{fi}$) is the spin operator 
for a conduction electron (spin-$\frac{1}{2}$ localized $f$-electron). 
Here, $t$ is the nearest-neighbor hopping matrix in the conduction 
band (referred to as the $c$ band in the following), 
$J_K$ describes the on-site Kondo coupling, and
$J$ denotes the RKKY-exchange 
among the nearest-neighbor $f$-spins. 
We consider the three-dimensional simple cubic lattice 
with dispersion 
$\varepsilon_k = - 2 t (\cos k_x + \cos k_y + \cos k_z)$ 
and the RKKY interaction 
$J_q = J (\cos q_x + \cos q_y + \cos q_z)$. 
We focus on the Kondo insulating 
phase at half-filling, and take  half the band width 
as the energy unit. 
The magnetic field $B$ applied along the $z$ direction is
coupled to both of the $c$ and $f$ electrons.  We assume that
the $g$ factors of the $c$ and $f$ electrons are
the same, and set $g \mu_B = 1$. 
We use the Kondo coupling $J_K = 0.5$ in following discussions. 

\subsection{Extended-DMFT for the Kondo lattice model}

DMFT is a powerful framework to study 
strongly correlated electron systems.
Since its power was proved 
by providing the first unified scenario 
for the longstanding probrem of the Mott transition 
in the Hubbard model \cite{kotliar}, 
various correlated electron models 
have been studied by using DMFT 
\cite{dmft,jarrell,rozenberg,imai_pam,
schork,kotliar,mutou,pruschke,ryouta,rozenberg_thm,
held,momoi,imai_hubbard,florens,koga}. 
EDMFT is an extension of DMFT so as to allow dynamical treatment of
inter-site interactions, which can be dealt with only at the
mean field level within the standard DMFT framework. 
EDMFT has successfully applied to 
the Kondo lattice model \cite{zhu,burdin,burdin2}, 
the periodic Anderson model \cite{sun_qcp}, 
the extended Hubbard model \cite{sun_ehm}, and 
the $t$-$J$ model \cite{haule}, etc.

In EDMFT the original lattice model is mapped onto an effective 
impurity model with a bosonic bath in addition to 
a fermionic bath that is included in the standard DMFT. 
The bosonic effective bath describes a fluctuating magnetic field 
and allows dynamical treatment of inter-site spin 
correlations. 
This approximation becomes exact in the limit of infinite 
dimensions, $d \to \infty$, 
provided that $t$ and $J$ are scaled as 
$t/\sqrt{d}$ and $J/\sqrt{d}$ respectively. 
It may give a good approximation even in three dimensions. 
We assume the paramagnetic Kondo insulating phase 
without long-range order. 
In this case, the Kondo lattice model (1) 
can be mapped onto the following 
Bose-Fermi Kondo Hamiltonian \cite{smith} 
as an effective impurity model: 
\begin{eqnarray}
H_\mathrm{EDMFT} &=& H_\mathrm{loc} 
  + H_\mathrm{bath} + H_\mathrm{mix} , \\
H_\mathrm{loc} &=&
  - \mu \sum_\sigma c^\dag_{\sigma} c_{\sigma}
  + J_K \mathbf{S}^c \cdot \mathbf{S}^f
  - B S^c_z - \tilde{B} S^f_z , \\
H_\mathrm{bath} &=& 
    \sum_{k,\sigma} E_{k\sigma} a^\dag_{k\sigma} a_{k\sigma}
  + \sum_{q,\alpha} \omega_{q\alpha} 
    h^\dag_{q\alpha} h_{q\alpha} , \\
H_\mathrm{mix} &=& 
    V \sum_{k,\sigma} \left [ 
    c^\dag_{\sigma} a_{k\sigma} 
  + a^\dag_{k\sigma} c_{\sigma} \right ] 
  + I \sum_{q,\alpha} S^f_\alpha  \left [
    h_{q\alpha} + h_{-q\alpha} \right ] .
\end{eqnarray}
Here, $H_\mathrm{loc}$ is the local part of 
the effective Hamiltonian at the site chosen as an `impurity', 
which includes the Hartree contribution of 
the RKKY-interaction among the $f$-spins, 
$\tilde{B} = B - 2J_{q=[0,0,0]} \langle S^f_z \rangle S^f_z$. 
The effective baths are described as $H_\mathrm{bath}$, 
where $a^\dag_{k\sigma}$ creates a fermionic bath 
with the dispersion $E_{k\sigma}$, and
$h_{q\alpha}$ $(\alpha = x,y,z)$ creates a bosonic bath 
with the dispersion $\omega_{q\alpha}$. The latter incorporates 
the effect of a fluctuating magnetic field. 
The `impurity' part and the effective baths 
are mixed by $H_\mathrm{mix}$, where the electron (spin) 
of the `impurity' couples to a fermionic (bosonic) bath
via $V$ ($I$). 
The parameters for the  effective baths, $E_{k\sigma}$, 
$\omega_{q\alpha}$, $V$ and $I$
are determined self-consistently by equating the Green's 
function and the spin susceptibility obtained 
in the effective impurity model with 
those in the original lattice model. 
The local Green's function for $c$ electrons and 
the susceptibility for $f$ moments are defined as 
$
G_{c\sigma} \left ( \tau \right ) = 
  - \left \langle T_{\tau} c_{i\sigma} (\tau) 
  c_{i\sigma}^{\dag}(0) \right \rangle
$
and 
$
\chi_{f\alpha} \left ( \tau \right ) = 
  \left \langle T_{\tau} S_{i\alpha}^{f}(\tau) 
  S_{i\alpha}^{f}(0) \right \rangle,
$
 which are calculated 
by solving the impurity model. 
We solve the effective Hamiltonian $H_\mathrm{EDMFT}$ 
by means of NCA \cite{keiter,bickers,schork} 
and calculate $G^{imp}_{c\sigma}$ and 
$\chi^{imp}_{f\alpha}$ \cite{haule}. 
The Green's function and the susceptibility satisfy the 
following equations, 
\begin{eqnarray}
G_{c\sigma}^{imp} (\omega) &=& 
  \left [ \omega + \mu + \frac{1}{2}\sigma B
  - G_{a\sigma} (\omega) 
  - \Sigma_\sigma (\omega) \right ]^{-1}, \\
\chi_{f\alpha}^{imp} (\omega) &=&
  \left [ G_{h\alpha} (\omega) + M_\alpha (\omega) \right ] ^{-1} ,
\end{eqnarray}
where $G_{a\sigma} (\omega)$ and $G_{h\alpha} (\omega)$ describe 
the effective fermionic and bosonic bath: 
\begin{eqnarray}
G_{a\sigma} (\omega) &=& 
  \sum_k \frac{V^2}{\omega - E_{k\sigma}} , \\
G_{h\alpha} (\omega) &=& \sum_q \left [ 
  \frac{I^2}{\omega - \omega_{q\alpha}} 
  - \frac{I^2}{\omega + \omega_{q\alpha}} \right ] .
\end{eqnarray}
We can obtain the momentum-independent 
self-energy $\Sigma_\sigma$ and 
$M_\alpha$ with these equations. 

The local Green's function and susceptibility
in the lattice system are 
obtained  in terms of the self-energy 
$\Sigma_\sigma$ and $M_\alpha$
by using the relations 
\begin{eqnarray}
G_{c\sigma}^{loc} (\omega) &=& \int d\varepsilon 
  \frac{\rho_t (\varepsilon)}
  {\omega + \mu + \frac{1}{2} \sigma B
  -\varepsilon - \Sigma_\sigma ( \omega )} , \\
\chi_{f\alpha}^{loc} (\omega) &=& \int d\varepsilon 
  \frac{\rho_J(\varepsilon)}
  {M_\alpha ( \omega ) + \varepsilon}, 
\end{eqnarray}
where $\rho_t (\varepsilon)$ and $\rho_J (\varepsilon)$ 
are the density of states for the free $c$-electrons and 
bosons representing spin fluctuations,
which are defined as 
$
\rho_t (\varepsilon) = \frac{1}{N} 
  \sum_k \delta (\varepsilon - \varepsilon_k)
$
and 
$
\rho_J (\varepsilon) = \frac{1}{N} 
  \sum_q \delta (\varepsilon - J_q)
$. 
Equations (6)-(11) complete the self-consistent loop. 

\section{Results}

\begin{figure}[hbt]
\begin{minipage}[t]{\minitwocolumn}
\begin{center}
\includegraphics[clip,trim=2.5cm 9.5cm 3cm 3cm,
                 width=\figsmall]{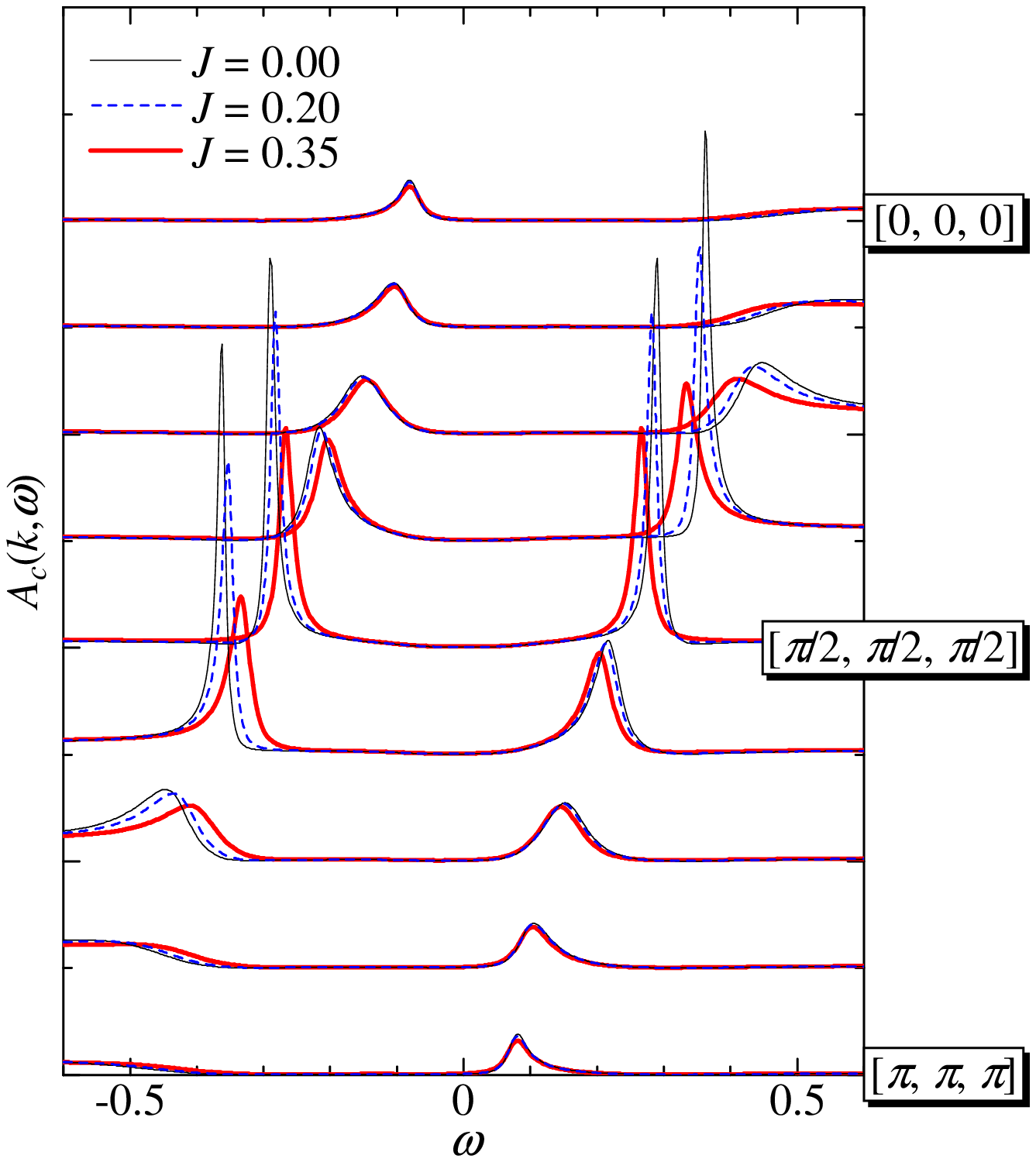}
\end{center}
\caption{The momentum resolved 
one-particle spectra $A_{c} (k, \omega)$ 
at $J_K=0.5$, $T=0.05$ for different values 
of the RKKY-interaction $J$. 
The data are plotted along the 
[1,1,1] direction in the $k$-space.}
\label{spectrum1}
\end{minipage}
\begin{minipage}[t]{\minitwocolumn}
\begin{center}
\includegraphics[clip,trim=2.5cm 9.5cm 3cm 3cm,
                width=\figsmall]{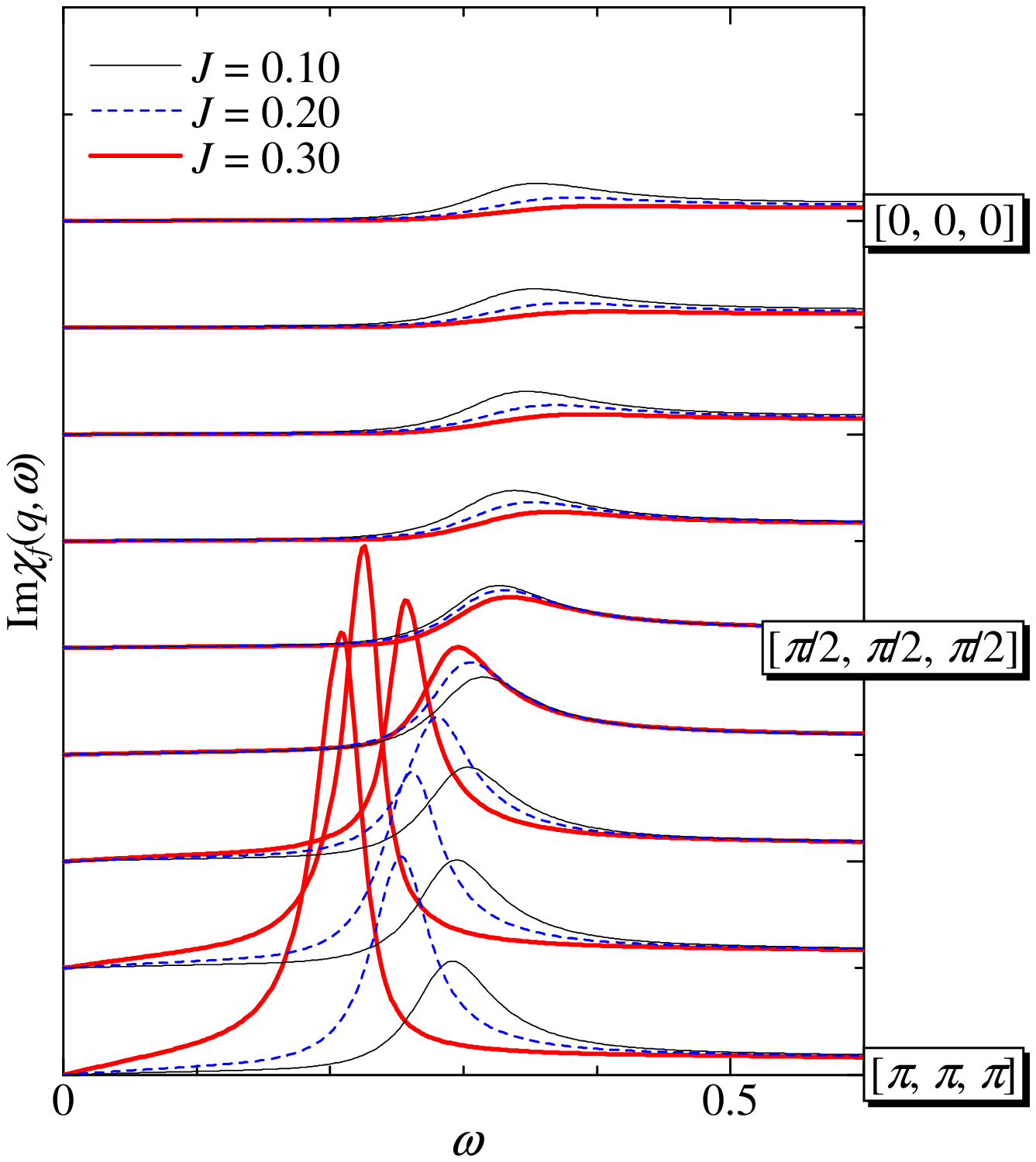}
\end{center}
\caption{The
dynamical spin susceptibility,  
$\mathrm{Im}\chi_{f} (q, \omega)$,  
at $J_K=0.5$ and $T=0.05$
for different values of the RKKY-interaction $J$.}
\label{spin1}
\end{minipage}
\end{figure}

We first discuss how the RKKY-exchange among $f$-spins 
renormalizes the one-particle spectrum and the spin dynamics
at $B=0$. 
One-particle Green's function $G_{c\sigma} (k, \omega)$  for 
conduction electrons is 
calculated with the self-energy $\Sigma_\sigma ( \omega )$, 
\begin{equation}
G_{c\sigma} (k, \omega) = \frac{1}{\omega + \mu 
  + \frac{1}{2} \sigma B -\varepsilon_k 
  - \Sigma_\sigma ( \omega )} .
\end{equation}
In Fig. \ref{spectrum1}, we plot the one-particle spectral function
$A_{c}(k,\omega)= -\mathrm{Im}G_{c\sigma} (k, \omega)/\pi$ 
for different values of the RKKY-interaction $J$ 
at a given temperature $T=0.05$. 
In the absence of $J$, 
the spectrum acquires an insulating gap 
resulting from the Kondo coupling $J_K$,
 since we
deal with  the half-filled band here. 
The size of the gap $\Delta$ can be defined as the 
energy difference between 
two peaks with the lowest excitation energy, which are located
at $k=[0,0,0]$ and $k=[\pi,\pi,\pi]$.
The gap is almost unchanged upon introducing the RKKY-interaction.
This result partly comes from the NCA approximation, which
has a tendency to overestimate the strength of Kondo singlets. 
We see, however, that the sharp peaks located around 
$k=[\pi /2,\pi /2,\pi /2]$  are suppressed 
and shifted to lower frequencies as the RKKY-interaction increases, 
signaling that the local singlet state gradually gets unstable 
in the presence of the RKKY-interaction. 

The effects of the RKKY-interaction are observed more clearly 
in spin excitations. 
We calculate the dynamical spin  susceptibility 
$\chi_{f\alpha} (q, \omega)$ with two-particle self-energy
$M_\alpha ( \omega )$, 
\begin{equation}
\chi_{f\alpha} (q, \omega) = 
  \frac{1}{M_\alpha ( \omega ) + J_q}. 
\end{equation}
In Fig. \ref{spin1} we plot the imaginary part of the dynamical 
susceptibility $\mathrm{Im}\chi_{f} (q, \omega)$
for different values of $J$. 
The dominant structure in $\mathrm{Im}\chi_{f} (q, \omega)$ 
is the low-frequency peak featuring spin-triplet excitations. 
As the RKKY-interaction increases, the spin-triplet peak at
$q=[\pi,\pi,\pi]$ is strongly enhanced and 
the position of the peak shifts to lower frequencies, 
indicating that the system gradually
approaches the antiferromagnetic instability. 

\begin{figure}[b]
\begin{minipage}[t]{\minitwocolumn}
\begin{center}
\includegraphics[clip,width=\figsmall]{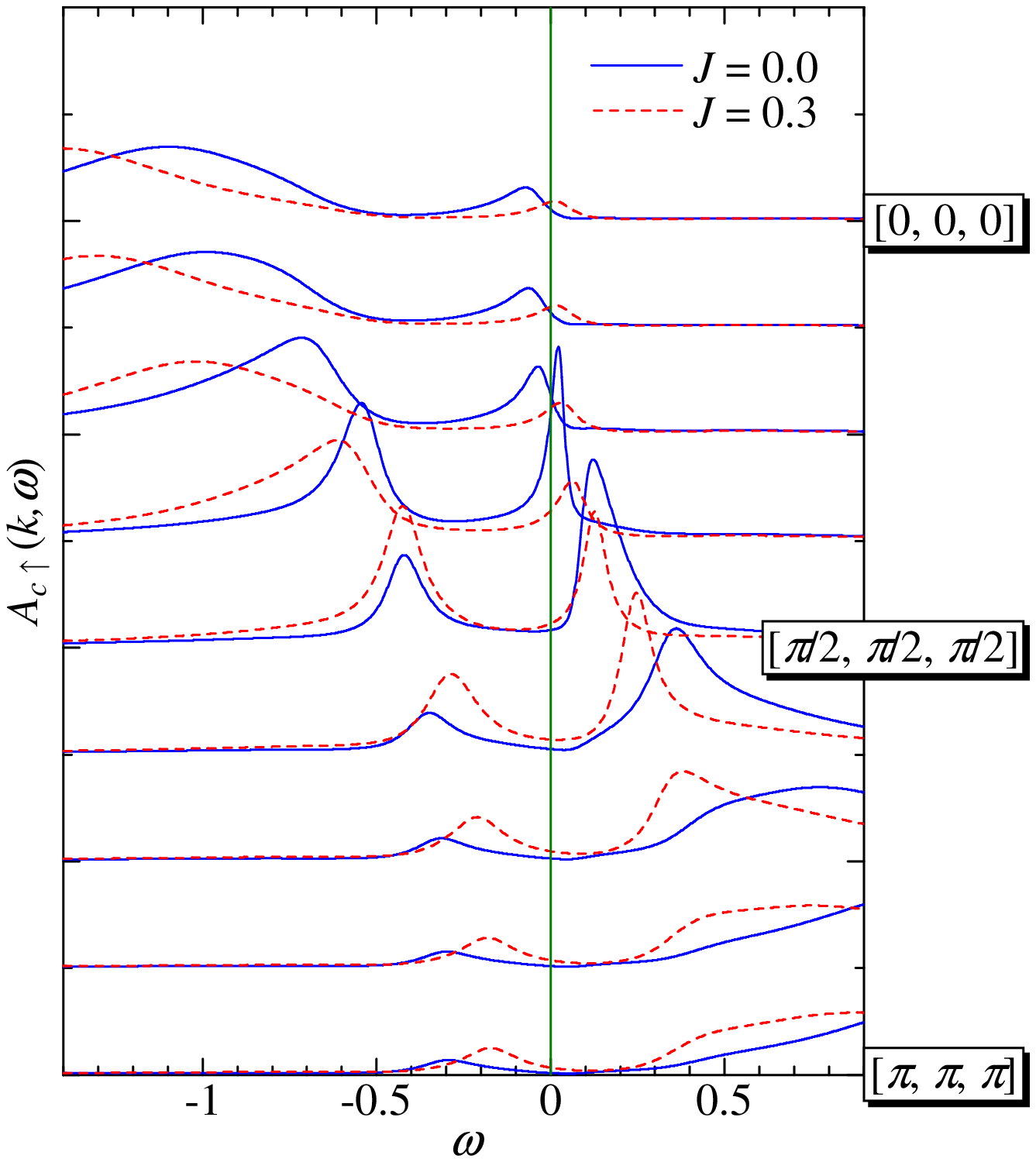}
\end{center}
\caption{The
one-particle spectral function  for up-spin 
(majority-spin) electrons, $A_{c\uparrow} (k, \omega)$,
at the magnetic field $B=0.3$:
$J_K=0.5$, $T=0.05$.
}
\label{spectrum2}
\end{minipage}
\begin{minipage}[t]{\minitwocolumn}
\begin{center}
\includegraphics[clip,width=0.7\minitwocolumn]{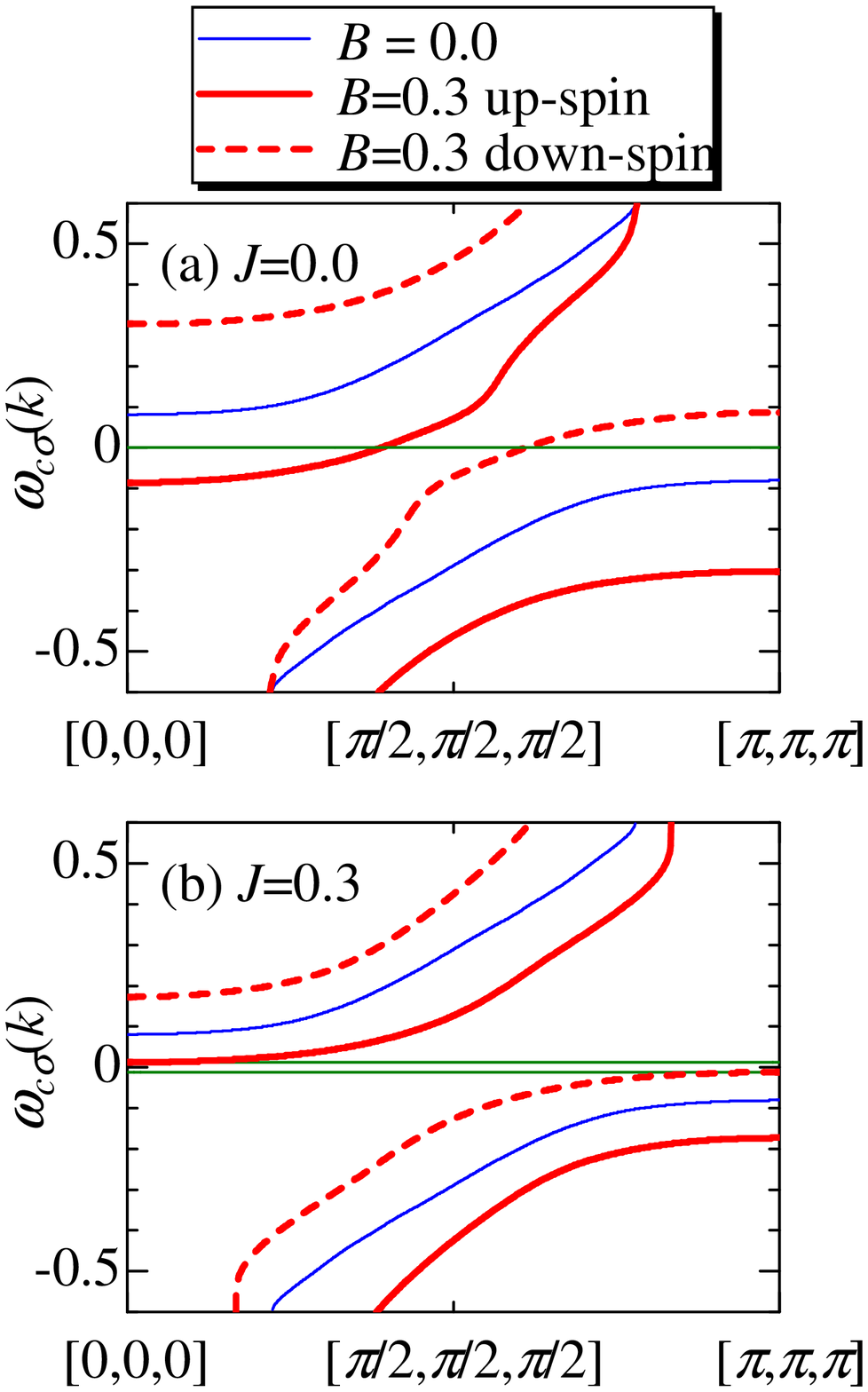}
\end{center}
\caption{
The quasi-particle dispersion $\omega_{c\sigma}(k)$
for (a) $J=0.0$  and (b) $J=0.3$. 
The dispersion is obtained by tracing the peak-position 
of the one-particle spectra $A_{c} (k, \omega)$. 
}
\label{dispersion}
\end{minipage}
\end{figure}

We now discuss how the effects of the RKKY-interaction 
are modified  in the presence of  a magnetic field. 
In Fig. \ref{spectrum2} the one-particle spectra for  up-spin 
electrons, 
$A_{c\uparrow} (k, \omega)$, are shown  at $B=0.3$. 
We recall that the Zeeman splitting due to the magnetic 
field pushes $A_{c\uparrow} (k, \omega)$ down to the negative-frequency 
side, so that the  peak position 
 at $k=[0,0,0]$ is shifted below the Fermi level, 
and thus the insulating gap is closed. This is indeed observed in 
the results for $J=0$ in Fig. \ref{spectrum2}.
Upon introduction of $J$, there 
appear two notable features. First, the effective Zeeman shift is 
increased in the presence of $J$, which is seen in the 
high frequency region of Fig. \ref{spectrum2}. This is caused by 
the competition between the Kondo and RKKY couplings. For $J=0$,
the Kondo-singlet formation suppresses the effect of the
Zeeman shift. Since the RKKY interaction enhances spin fluctuations of
$f$-electrons, it makes the Kondo effect weaker, thus
reviving the large Zeeman shift.  

\begin{figure}[bth]
\begin{center}
\includegraphics[clip,width=\figwidth]{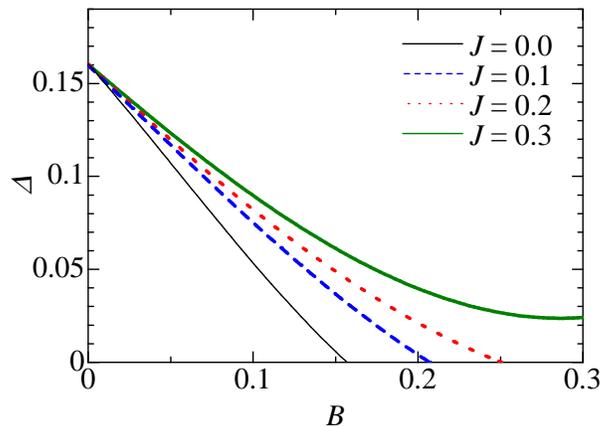}
\end{center}
\caption{
The magnetic field dependence 
of the gap for different values of $J$. 
}
\label{gap}
\end{figure}

Another important feature is the characteristic low-energy 
behavior. For example, in the
low-energy region for $J=0.3$, the dispersion
is highly renormalized to form a flat band
around $\omega\sim 0$, which cannot be explained in terms of the
effective Zeeman shift mentioned above.
Such behavior is seen more clearly in the quasi-particle dispersion 
of $c$-electrons, $\omega_{c\sigma}(k)$, plotted in Fig. \ref{dispersion}.
This remarkable effect is regarded as
the mass enhancement caused by critical $f$-spin fluctuations.
In fact, the system at  $J=0.3$ is very close to 
the quantum phase transition, as discussed momentarily below.
 Conduction electrons scattered by 
such critical $f$-spin fluctuations get heavier, resulting in 
the flat dispersion. The corresponding gap is shown in Fig. \ref{gap}
as a  function of the field.
In the absence of the RKKY-interaction, the gap is reduced 
linearly and closed at the field corresponding to the gap size
$B \sim 0.16$. 
As the RKKY-interaction is increased, 
the gap can persist up to larger fields
reflecting the renormalization effect mentioned above.

\begin{figure}[t]
\begin{minipage}[t]{\minitwocolumn}
\begin{center}
\includegraphics[clip,trim=3cm 6.5cm 1.5cm 4.5cm,
                 width=\figsmall]{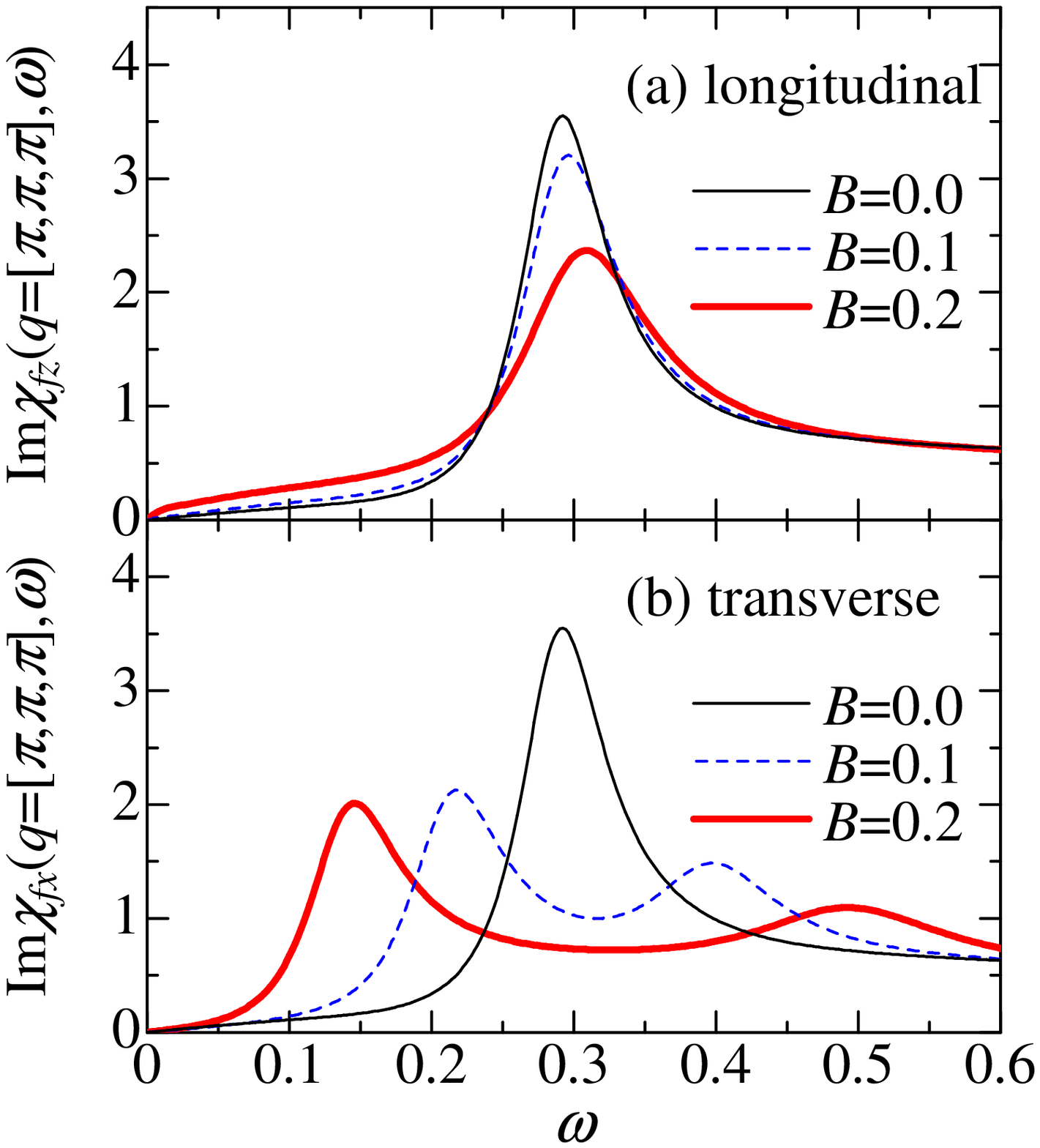}
\end{center}
\caption{
The imaginary part of the dynamical susceptibility
$\mathrm{Im}\chi_{f\alpha} (q=[\pi,\pi,\pi], \omega)$:
(a) $z$ direction and (b) $x$ direction 
for different magnetic fields at $J=0.1$. 
}
\label{fig_a1}
\end{minipage}
\begin{minipage}[t]{\minitwocolumn}
\begin{center}
\includegraphics[clip,trim=3cm 6.5cm 1.5cm 4.5cm,
                 width=\figsmall]{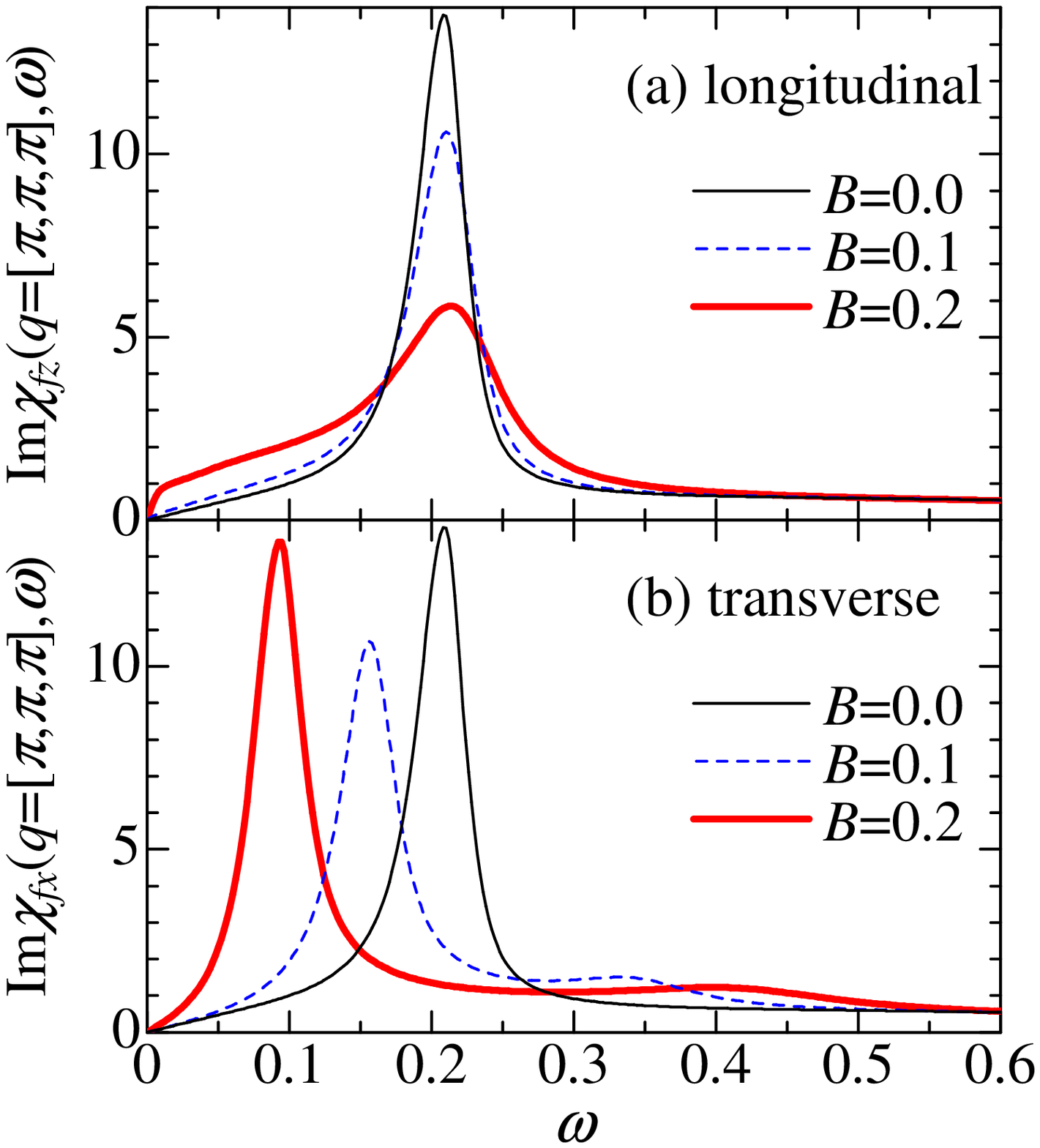}
\end{center}
\caption{
The same plots of the dynamical susceptibility as in Fig.
\ref{fig_a1}. The RKKY interaction is chosen as $J=0.3$. 
}
\label{fig_a2}
\end{minipage}
\end{figure}

\begin{figure}[t]
\begin{center}
\includegraphics[clip,width=\figwidth]{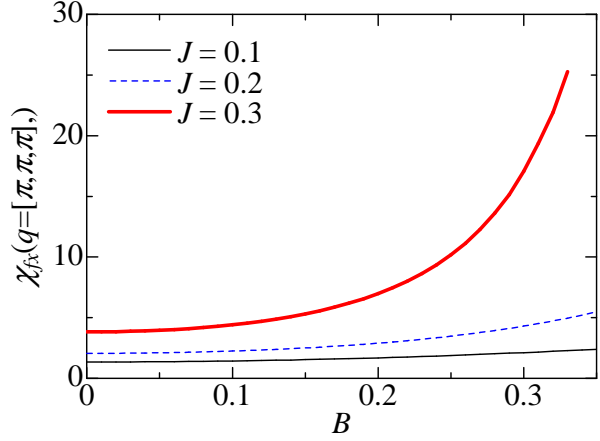}
\end{center}
\caption{
The transverse spin susceptibility 
in the $x$ direction $\chi_{fx}(q=[\pi, \pi, \pi])$ 
as a function of the magnetic field $B$ for different 
strength of the RKKY-interaction $J$.
}
\label{sus}
\end{figure}

In order to see  how $f$-spin fluctuations behave in magnetic fields,
we plot the field dependence of 
the imaginary part of the dynamical susceptibility 
$\mathrm{Im}\chi_{f\alpha} (q=[\pi,\pi,\pi], \omega)$ 
$(\alpha = x, z)$  in Figs. \ref{fig_a1} and \ref{fig_a2}.
Comparing the susceptibility for $J=0.1$ and 0.3 at $B=0$,
one can see that $[\pi,\pi,\pi]$ triplet modes are softened in 
 the presence of the RKKY interaction.
As the magnetic field is introduced,
the longitudinal ($z$-component) and the transverse
($x$-component) susceptibilities exhibit different properties.
Since $\mathrm{Im}\chi_{fz} (q=[\pi,\pi,\pi], \omega)$ 
excites an $S_z=0$ triplet mode,  its
 peak position is mostly unchanged even at finite fields, 
as is seen both for the $J=0.1$ and $0.3$ cases.
On the other hand, the transverse susceptibility 
$\mathrm{Im}\chi_{fx} (q=[\pi,\pi,\pi], \omega)$
shows a splitting in the spectrum, corresponding to $S_z=\pm 1$
triplet modes. The peak in the higher-energy side is smeared
considerably by the life-time effect. It should be noticed 
here that the peak structure, which is shifted to lower frequencies,
gets sharper as $B$ increases, and this tendency becomes
more significant as $J$ becomes large.  
 For example, for the $J=0.3$ case, although weak magnetic fields ($B=0.1$) 
merely suppress spin fluctuations and thus reduce the spin-triplet peak, 
larger magnetic fields ($B=0.2$) increase
the peak intensity again, signaling that
the system approaches the antiferromagnetic instability.
To see the antiferromagnetic instability directly, 
we show the transverse static susceptibility of localized 
 $f$-spins, $\chi_{fx}(q=[\pi, \pi, \pi])$,
in Fig. \ref{sus}.  For smaller RKKY couplings ($J=0.1,0.2$)
there is no indication of the instability as the magnetic field increases.
We can see  the rapid increase of 
$\chi_{fx}(q=[\pi, \pi, \pi])$  for
$J=0.3$ reflecting the antiferromagnetic instability:
 the system is driven to the transverse antiferromagnetic
phase ($x$-$y$ plane) with finite values of 
$z$-components (canted order).

\begin{figure}[tbh]
\begin{flushright}
\includegraphics[clip,width=\figlarge]{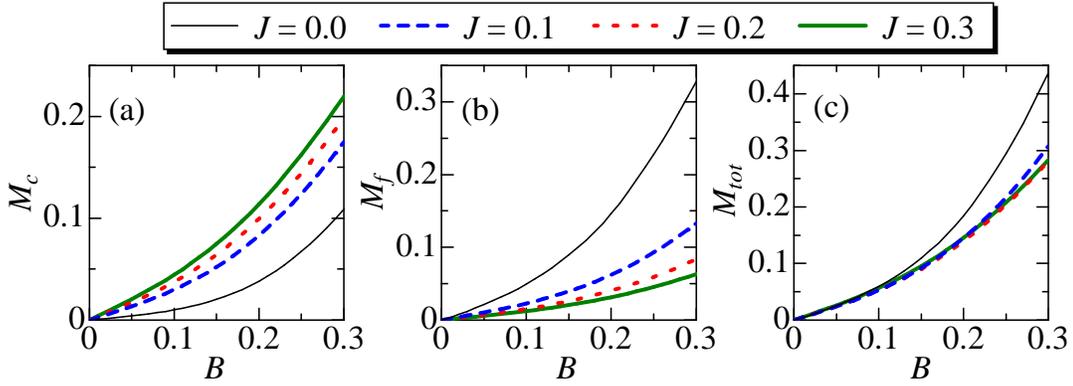}
\end{flushright}
\caption{
The magnetization process:(a)  $c$-electrons, (b) $f$-electrons 
and (c) the total contribution. 
$M_c$, $M_f$ and $M_{tot}$ are defined by 
$M_c=\langle S_c^z \rangle$, 
$M_c=\langle S_f^z \rangle$, and $M_{tot}=M_c+M_f$.
}
\label{magnet}
\end{figure}

\begin{figure}[tbh]
\begin{center}
\includegraphics[clip,width=\figwidth]{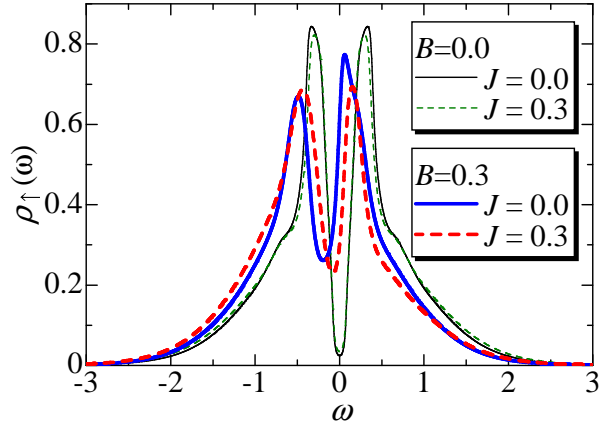}
\end{center}
\caption{The density of states for up-spin conduction electrons for
some different choices of the magnetic field $B$ and 
the RKKY-interaction $J$.
}
\label{DOS}
\end{figure}

The above characteristic properties 
also affect the static quantities
in magnetic fields.  For instance, we plot the magnetization curve in
Fig. \ref{magnet}. 
At $J=0$,  $M_f$ ($M_c$) has a relatively large (small) response to
the field. As the strength of the RKKY interaction increases, 
$M_f$ is suppressed whereas  $M_c$ is enhanced.
Although the sharp spin-gap behavior in the magnetization
 is somehow obscured for the parameters exploited here, 
the above magnetic-field dependence indeed comes from the 
competition between
the Kondo coupling and the RKKY interaction, as explained below.
 The increase of $J$ has a tendency to suppress
the Kondo singlet formation, so that the conduction electrons
start to polarize rather freely for finite $J$, 
increasing the magnetization. 
On the other hand, 
$M_f$ is generally suppressed with the increase of  $J$. 
Such behavior is consistent with the change of the density of states 
(DOS) shown in  Fig. \ref{DOS}.  Comparing the DOS between the cases of
 $J=0$ and 0.3, we see that the effective Zeeman shift 
in conduction electrons is somewhat enhanced for $J=0.3$,
which is clearly observed around $|\omega| \sim 1$. This shift
gives rise to the increment of $M_c$ as seen in Fig. \ref{magnet}.

Finally, we make a comment on the low-energy behavior of the
DOS in  Fig. \ref{DOS}. If we look at the DOS around $\omega \sim 0$,
the shift of the peak structure is smaller for $J=0.3$ 
than $J=0$ contrary to the effective Zeeman shift.
This is what we have claimed  as the mass renormalization effect due to 
critical transverse $f$-spin fluctuations in Fig. \ref{spectrum2}; it is 
a typical correlation effect  between conduction electrons and
$f$-spins, which is beyond the naive mean-field picture.

\section{Summary}

We have studied the KLM with the 
RKKY exchange interaction among local $f$-spins
in the presence of a magnetic field. 
For this purpose we have exploited an extended 
version of  DMFT combined with NCA. 
This approach enables us to treat inter-site spin 
correlations dropped in the DMFT framework.
By investigating the one-particle spectral function 
and the dynamical spin susceptibility, we have 
confirmed that the RKKY-interaction has a tendency to 
enhance field-induced magnetic instability to the 
transverse magnetic order of  $f$-spins, as should be expected.
We have clarified two characteristic features caused by
the competition of the Kondo 
 and RKKY interactions in magnetic fields.
One is a simple mean-field type effect: the RKKY
interaction enhances the Zeeman shift for conduction electrons, 
which modifies the profile of the one-particle spectrum 
at high frequencies as well as the magnetization curve. 
Another important feature is 
a renormalization effect observed in the low-energy region: 
the quasi-particle spectrum gets renormalized to form
heavy quasi-particles when the system is close to the phase transition point.
This effect appears as if the RKKY interaction 
protects the Kondo insulating gap in magnetic fields.
Anyway, this remarkable feature at low energies 
is a typical correlation effect beyond the mean field picture,
which characterizes the critical region of the field-induced
phase transition in the Kondo insulator.

Field-induced transitions to 
the transverse antiferromagnetic order should be observed
in heavy-fermion Kondo insulators. 
Unfortunately, there have been few examples of 
such transitions reported so far.
Recently, it has been experimentally found \cite{rotundu,sugawara} 
that a typical filled-skutterdite, $\mathrm{CeOs_4Sb_{12}}$, 
which may be classified as an Kondo insulator, 
exhibits a field-induced magnetic phase transition. 
This transition could be a transition to the transverse 
antiferromagnetic  phase discussed here, 
although more detailed study should be done since the above 
skutterdite has rather complicated electronic structure 
including orbital degrees of freedom. 
We think that a variety of compounds exhibiting 
such field-induced transitions may be found 
in the class of heavy fermion Kondo insulators.

\ack
The authors thank A. Koga and  H. Tsunetsugu for valuable discussions.  
This work was partly supported by a Grant-in-Aid from the Ministry 
of Education, Science, Sports and Culture of Japan. 

\section*{References}

\end{document}